%% file: arxiv.tex
\documentclass[apj]{emulateapj}

\def\E{{\cal E}}
\def\F{{\cal F}}

\begin{document}

\title{Rings of Dark Matter in Collisions Between Clusters of Galaxies}

\author{J. A. ZuHone and D. Q. Lamb}
\affil{Department of Astronomy and Astrophysics, University of Chicago, Chicago, IL 60637}
\and
\author{P. M. Ricker}
\affil{Department of Astronomy, University of Illinois at Urbana-Champaign, Urbana, IL, 61801}

\begin{abstract}
Several lines of evidence suggest that the galaxy cluster Cl0024+17, an apparently relaxed system, is actually a collision of two clusters, the interaction occurring along our line of sight. Recent lensing observations suggest the presence of a ring-like dark matter structure, which has been interpreted as the result of such a collision. In this paper we present $N$-body simulations of cluster collisions along the line of sight to investigate the detectability of such features. We use realistic dark matter density profiles as determined from cosmological simulations. Our simulations show a ``shoulder'' in the dark matter distribution after the collision, but no ring feature even when the initial particle velocity distribution is highly tangentially anisotropic ($\sigma_\theta/\sigma_r >> 1$).  Only when the initial particle velocity distribution is circular do our simulations show such a feature.  Even modestly anisotropic velocity distributions are inconsistent with the halo velocity distributions seen in cosmological simulations, and would require highly fine-tuned initial conditions.  Our investigation leaves us without an explanation for the dark matter ring-like feature in Cl 0024+17 suggested by lensing observations.
\end{abstract}


\section{Introduction}

Clusters of galaxies have proven to be interesting laboratories for studying the dynamics of the different kinds of matter in the universe. Early measurements of galaxy velocity dispersions in clusters demonstrated that most of the mass of galaxy clusters was in the form of a non-luminous component \citep{zwy37}. Modern cosmological constraints (e.g. Turner 2000) suggest that this non-luminous material must be mostly non-baryonic. In addition, X-ray observations of clusters of galaxies reveal that most of the baryonic mass in clusters is in the form of a hot, diffuse, X-ray emitting gas, the intracluster medium (ICM). The presence of these different types of matter (stellar, gaseous, and dark) in these systems provides not only insights into the nature of the clusters themselves but also the nature of the different components of matter and their dynamics. 

Galaxy clusters are themselves the product of many mergers between galaxies, galaxy groups, and smaller clusters of galaxies \citep{dav85}. Observations indicate that this process is still ongoing in many systems. A famous example is the so-called ``Bullet Cluster'' (1E 0657-56), a merging system in which X-ray and weak lensing observations demonstrate a clear separation between a collisionless dark matter component and the intracluster medium \citep{mar02, clo06}.  

A system that has also garnered attention recently is Cl 0024+17, a cluster at $z = 0.395$ with weak and strong lensing. Early attempts at reconstructing the mass profile of this system using lensing \citep{tys98, bro00, com06} suggested a conflict with the predictions of the standard cold dark matter (CDM) model due to the flattening of the density profile in the inner regions of the cluster. Cosmological simulations assuming CDM indicate that galaxy cluster mass profiles should exhibit a logarithmic slope in the inner regions (e.g. Navarro, Frenk, \& White 1997). This apparent discrepancy (and others) led to suggestions that the CDM paradigm would need to be modified \citep{spr00,hog00,moo00}, for example to include self-interaction of the dark matter. 

\citet{czo01, czo02} demonstrated that the redshift distribution of the cluster galaxies in Cl~0024+17 is bimodal and suggested that the system is composed of two clusters undergoing a collision along the line of sight. They also performed a simulation demonstrating that such a scenario reproduces not only the bimodal redshift distribution but also the flattening of the central density profile. Observations of the cluster by Chandra \citep{ota04} and XMM-Newton \citep{zhg05} revealed that the surface brightness profile is better fit by a superposition of two ICM models rather than one. They suggested on the basis of the isothermal temperature profile that the system had returned to equilibrium after the collision and that consequently the encounter must have occured several Gyr ago. 

Recently, a weak lensing analysis presented by \citet{jee07} revealed a ring-like structure in the projected matter distribution. They proposed that dark matter from the cores of the clusters had been disrupted and ejected from the systems by the collision, and they demonstrated that such features could be reproduced in a simulation of a collision of two pure dark matter halos. On the basis of this result they suggested the current state of the system is $\sim$1-2 Gyr past the pericentric passage of the cluster cores.

We wish to shed additional light on the formation of ring structures within the dark matter distribution. ``Ring galaxies'' provide another astrophysical situation in which ``particle'' rings are produced due to collisions (e.g., the Cartwheel Galaxy). We look to the previous $N$-body simulations of these phenomena to guide our understanding of what might lead to such a feature in the dark matter in clusters of galaxies. Two major differences exist between the dark matter particles in galaxy clusters and star ``particles'' in disk galaxies. In the case of disk galaxies, the stars are concentrated in a disk, and the velocities are essentially circular. In galaxy clusters, dark matter particles form a roughly spheroidal distribution and have isotropic to radially anisotropic velocity dispersions. To investigate the effects of the second of these differences we perform a set of pure dark matter ($N$-body only) simulations with a physically motivated dark matter profile and variations on the velocity distribution of the dark matter particles. 

Throughout this paper we assume a flat $\Lambda$CDM cosmology with $h = 0.7$ and $\Omega_{\rm m} = 0.3$. 

\section{Simulations}
\label{Sec:simulations}

\subsection{Method}

We performed our simulations using FLASH, a parallel hydrodynamics/$N$-body astrophysical simulation code developed at the Center for Astrophysical Thermonuclear Flashes at the University of Chicago \citep{fry00}. FLASH uses adaptive mesh refinement (AMR), a technique that places higher resolution elements of the grid only where they are needed. AMR allows us to achieve high resolution these regions without having to fully resolve the whole grid. FLASH includes an $N$-body module which uses the particle-mesh method to solve for the forces on gravitating particles. The gravitational potential is computed using a multigrid solver included with FLASH \citep{ric08}.
 
\subsection{Initial Conditions}

For ease in setting up the particle distribution functions, we use a Hernquist profile \citep{her90}:

\begin{equation}
\rho_{\rm DM}(r) = \rho_s\frac{1}{r/a(1+r/a)^3}, 
\end{equation}

where $\rho_s = M_0/(2{\pi}a^3)$ is the scale density of the profile. The Hernquist mass profile that converges to $M_0$ as $r \rightarrow \infty$:

\begin{equation}
M(r) = M_0{\frac{r^2}{(r+a)^2}}
\end{equation}

This form of the dark matter density profile is chosen because of the mathematical simplicity of the corresponding distribution functions (as will be shown below) and its resemblance to the NFW profile in that as $r \rightarrow 0$, $\rho(r) \propto r^{-1}$. 


\input tab1.tex

To determine the effects of a varying velocity anisotropy on the dark matter features in the simulation, we allow the 3D velocity dispersion of the particles to vary from the isotropic form $\sigma_r = \sigma_\theta = \sigma_\phi$. We assume $\sigma_\theta = \sigma_\phi$ and parametrize the anisotropy using $\beta \equiv 1 - {\sigma_\theta^2 \over \sigma_r^2}$, which is taken to be constant over the entire radial range. Table \ref{tab1} shows the values of $\beta$ that were used for each simulation, and the corresponding ratio ${\sigma_\theta \over \sigma_r}$.

In order to initialize the particle velocities, we sample the particle distribution function (hereafter DF) directly. The DF $\F({\bf r,v})$ is assumed to obey the following relation:

\begin{equation}
\rho(r) = \int\F({\bf r,v})d{\bf v}
\end{equation}

The DF of any steady-state, spherically symmetric system has a dependence on the phase space coordinates that comes in only through the ``integrals of motion'' $\E$ and $L$, where $\E = \psi - \frac{1}{2}v^2$ is the relative energy and $L = rv_t$ is the angular momentum of a particle \citep{bin87}. In these expressions, $\psi(r) = -\phi(r)$ is the relative gravitational potential, and $v_t$ is the tangential velocity. 


\input tab2.tex

For a constant velocity anisotropy, the DF takes the specific form \citep{bin87}:

\begin{equation}
\F(\E, L) = L^{-2\beta}f_{\E}(\E)
\end{equation}

For a Hernquist profile, the energy-dependent part of the DF is \citep{bae02}

\begin{eqnarray}
\lefteqn{f_{\E}(\E) = \frac{2^{\beta}}{(2\pi)^{5/2}}\frac{{\Gamma}(5-2{\beta})}{{\Gamma}(1-{\beta}){\Gamma}(\frac{7}{2}-{\beta})}\E^{5/2-\beta}{}}
\nonumber\\ & & {}\times {_{2}F_{1}\left(5-2\beta,1-2\beta,\frac{7}{2}-\beta;\E\right)}
\end{eqnarray}

where $_{2}F_{1}$ is the hypergeometric function. For half-integer values of $\beta$ this function can be expressed in terms of rational functions. To ease our investigation of this parameter space we therefore choose values of $\beta$ from this set, specifically the values $\beta$ = 1/2, -3/2, -5/2, -17/2. We also include an isotropic setup ($\beta$ = 0) and a setup where all velocities are initially circular ($\beta = -\infty$). Given the mass density function and the distribution function, we derive initial positions and velocities for the particles using the acceptance-rejection method.

Finally, following \citet{czo02} and \citet{jee07} we assume a mass ratio of 2:1 for the clusters. In all of our simulations the clusters are initialized so that their centers are separated by the sum of their respective radii $R$ (3 Mpc), and they are given an initial relative speed $v_{\rm rel} = 3000$ km/s, which is the inferred relative velocity of the two components of Cl0024+17 seen in the redshift histograms of \citet{czo02}. In that study they showed that a initial velocity of $\sim$3000 km/s in their collision simulation reproduced this observed distribution of redshifts. This value for the velocity is approximately 2$v_{ff}$ for the clusters, where $v_{ff}$ is the free-fall velocity from infinity. Such an initially high relative velocity might be difficult to achieve in a $\Lambda$CDM universe \citep[see, e.g.][]{hay06}. However, our choice is motivated by our desire to match the relative inferred velocity of the cluster components of Cl~0024+17. The values of the halo parameters are given in Table \ref{tab2}.

We refine the adaptive mesh using the dark matter density to resolve the cores of the clusters and other overdensities. For our box size of 10$h^{-1}$ Mpc we achieve a minimum zone spacing of ${\Delta}x = 9.77h^{-1}$ kpc.


\begin{figure*}
\plotone{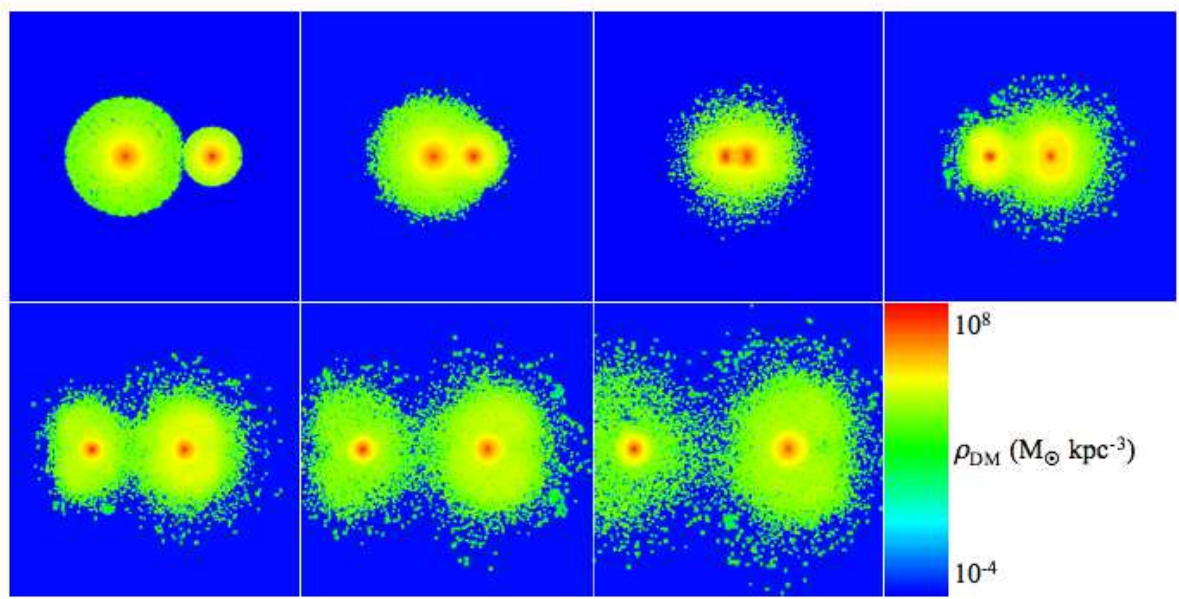}
\caption{Slices through the $z = 0$ coordinate plane of dark matter density at the epochs $t$ = 0.0, 0.5, 1.0, 1.5, 2.0, 2.5, and 3.0 Gyr for the $\beta = -5/2$ simulation. Each panel is 10 Mpc on a side.\label{fig1}}
\end{figure*}


\begin{figure*}
\includegraphics[width=2.0in]{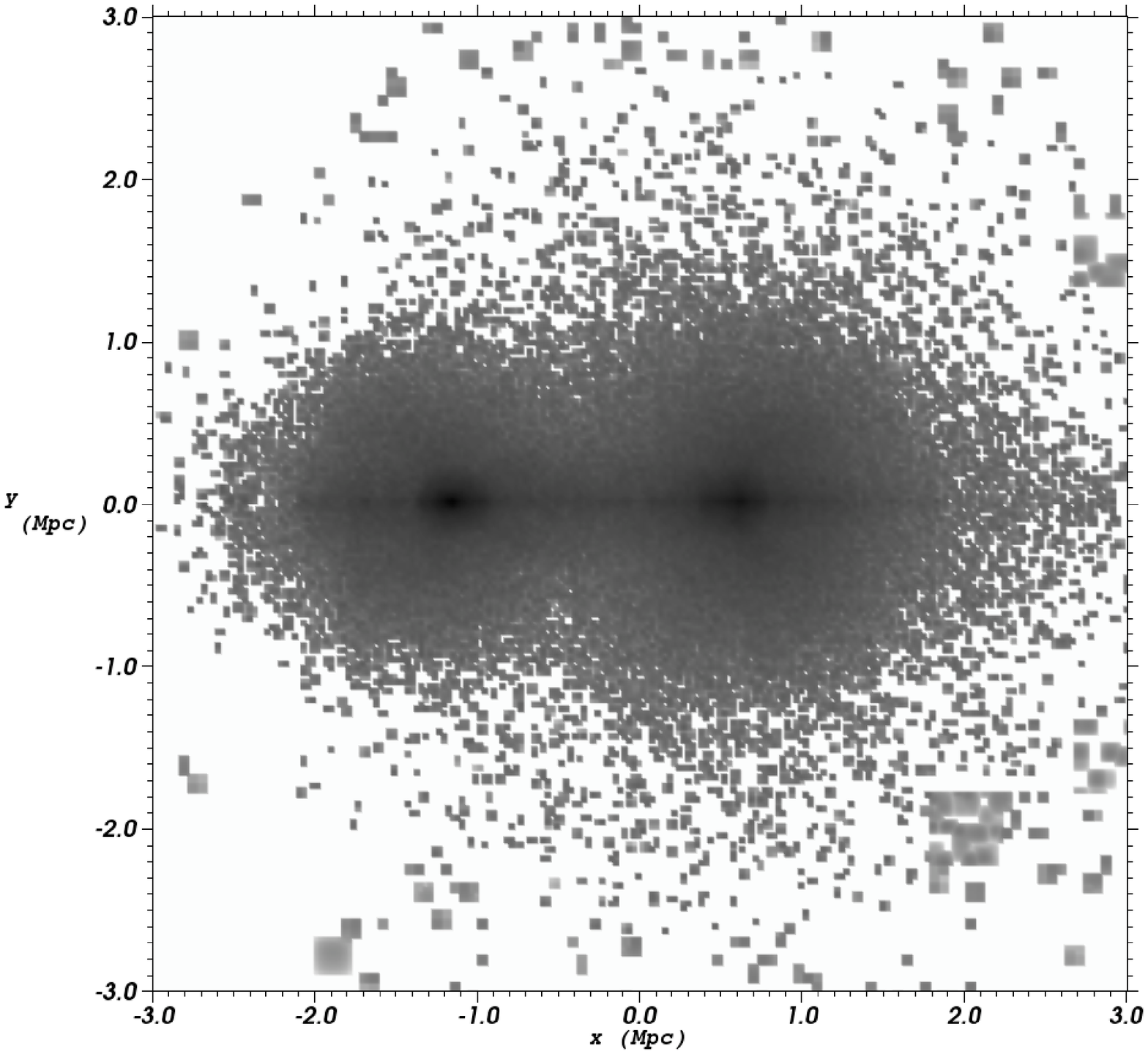}
\includegraphics[width=2.0in]{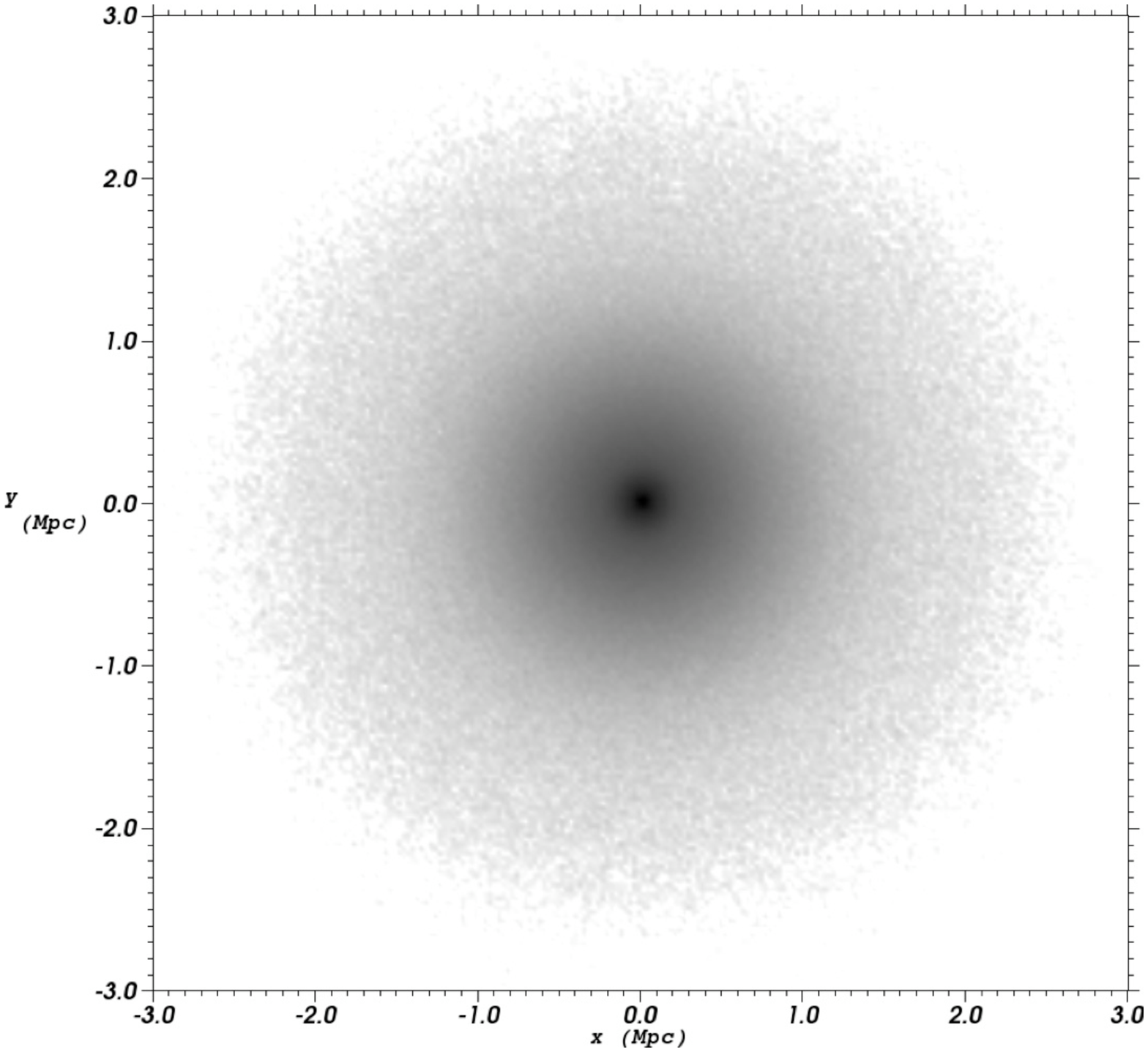}
\includegraphics[width=2.0in]{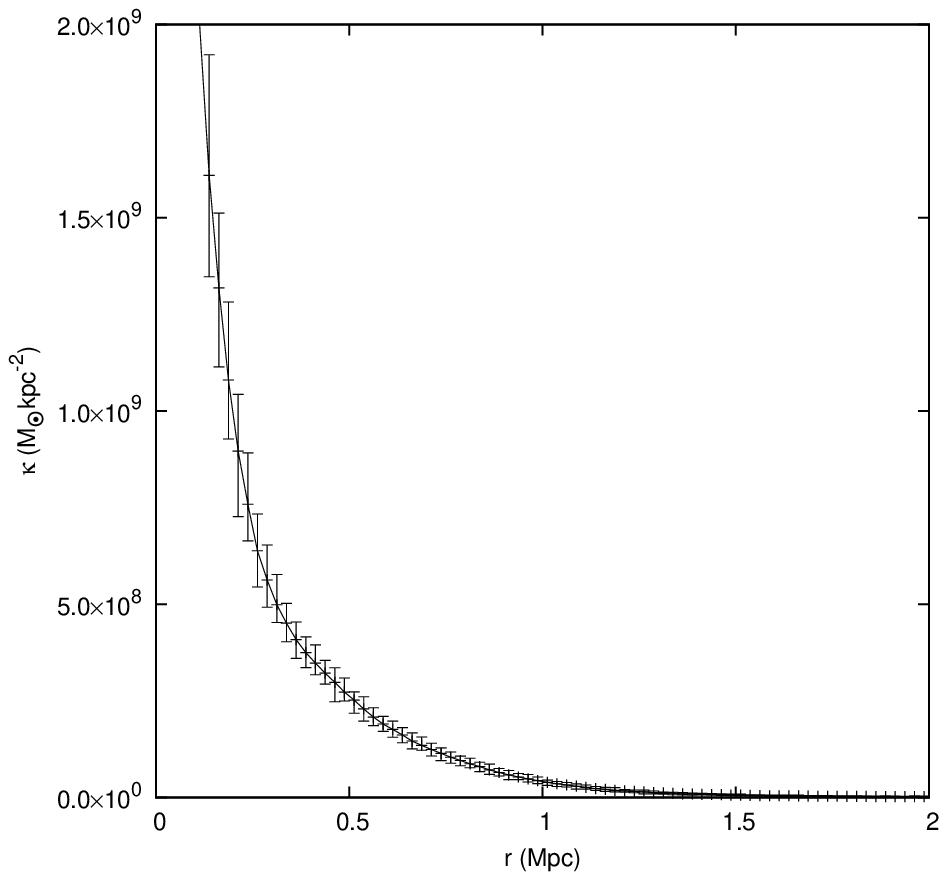}
\caption{$t$ = 1.5 Gyr snapshot of the $\beta$ = 1/2 simulation. Left: Slice through the $z = 0$ coordinate plane of dark matter density. Center: Dark matter density projected along the collision axis. Right: Projected density profile.\label{fig2}}
\end{figure*}

\section{Results}
\label{Sec:results}

For our set of simulations we have used the same cluster profiles, masses, initial separation, and relative velocity, but we have varied the anisotropy of the dark matter velocity dispersion. In each simulation, from $t$ = 0.0-0.9 Gyr the dark matter cores accelerate toward each other and pass through each other. The resulting sudden impulse on the dark matter particles boosts their energies and causes a significant fraction of the mass in the cores to be ejected from the centers. Figure \ref{fig1} shows slices through the $z = 0$ coordinate plane of the dark matter density in the $\beta = -5/2$ simulation at different epochs. After the pericentric passage the outer regions of both halos have expanded, with a trail of dark matter strung between the centers. It can be seen that the smaller halo is more disrupted, with a larger amount of mass redistributed to the outer portion of the halo. 

\begin{figure*}
\includegraphics[width=2.0in]{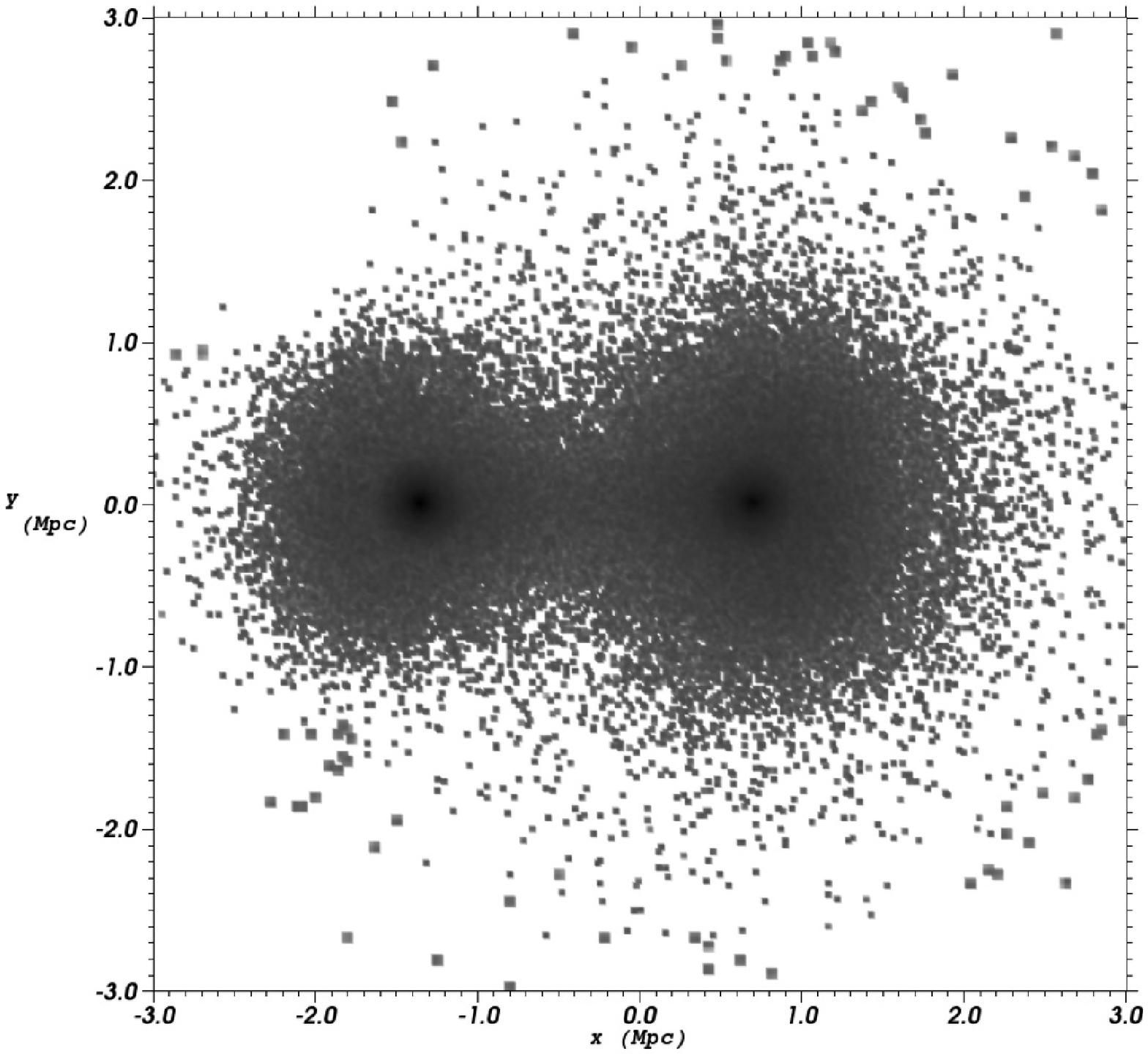}
\includegraphics[width=2.0in]{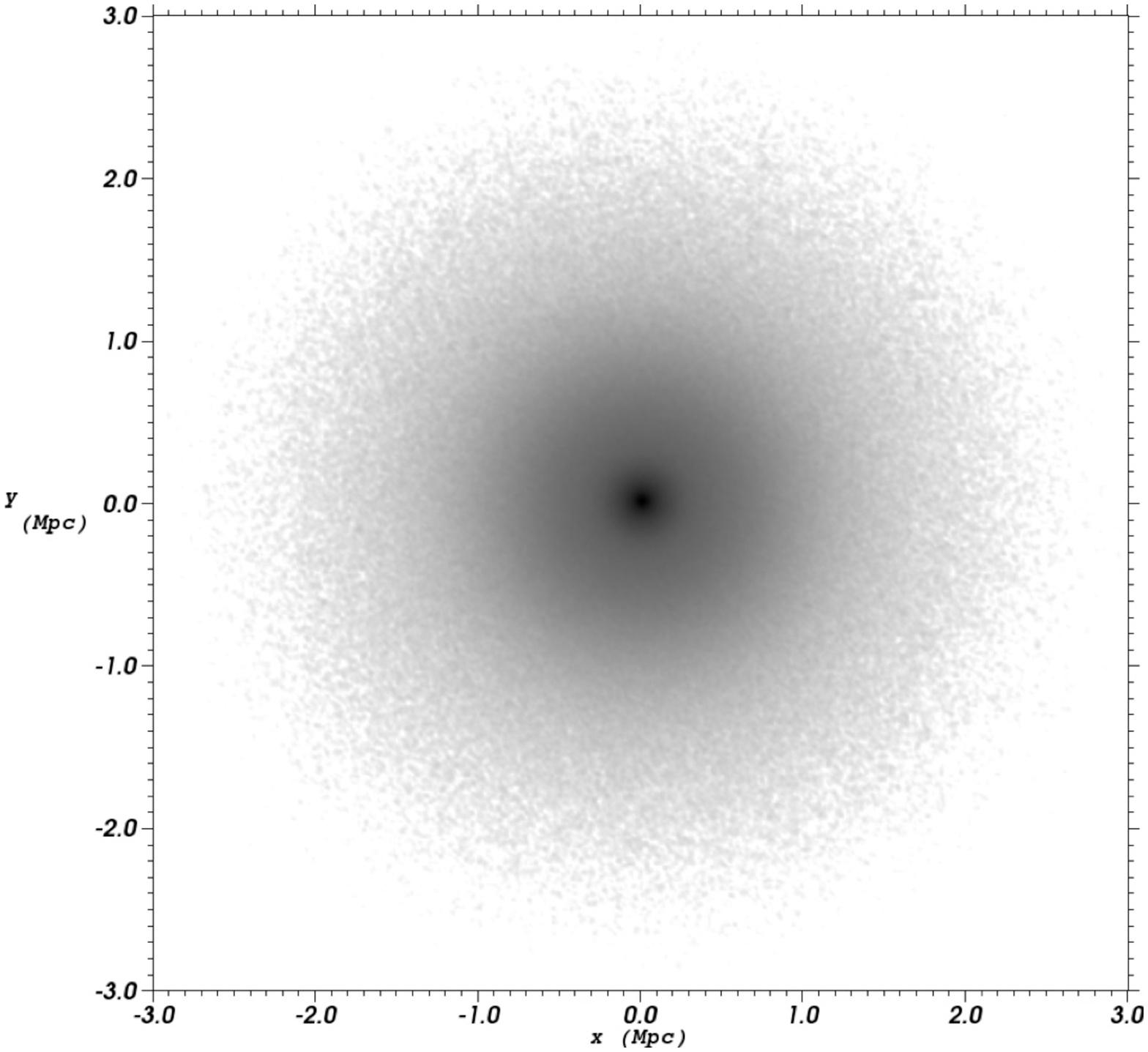}
\includegraphics[width=2.0in]{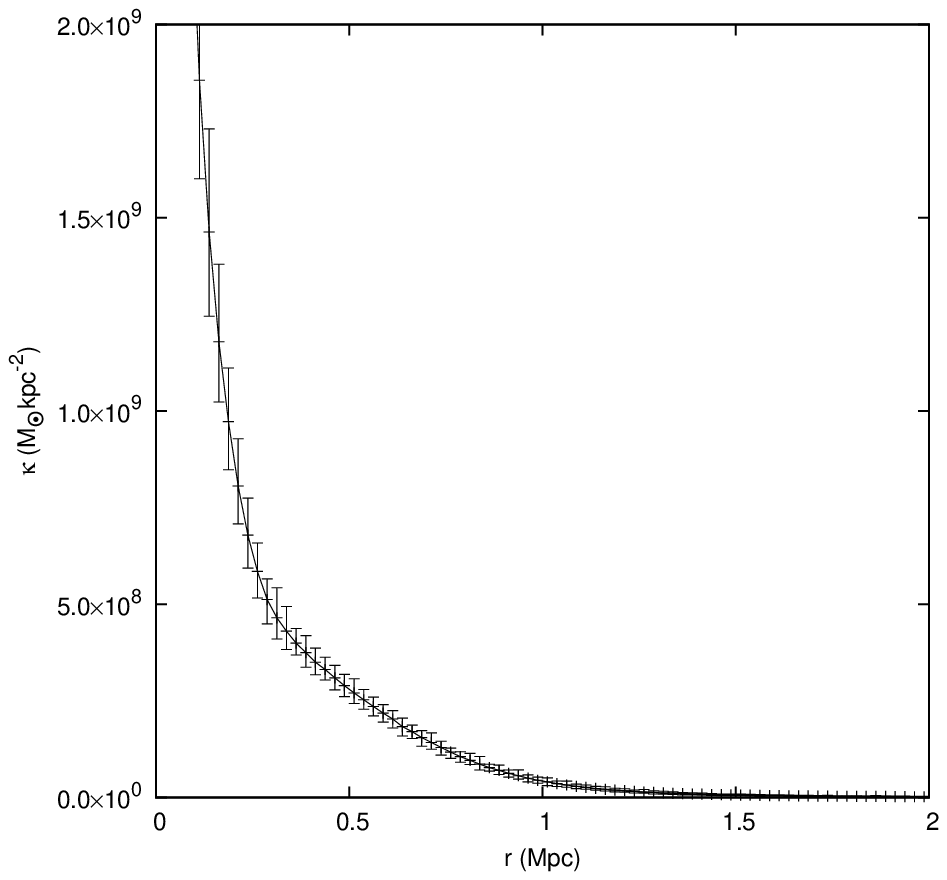}
\caption{$t$ = 1.5 Gyr snapshot of the $\beta$ = 0 simulation. Left: Slice through the $z = 0$ coordinate plane of dark matter density. Center: Dark matter density projected along the collision axis. Right: Projected density profile.\label{fig3}}
\end{figure*}

\begin{figure*}
\includegraphics[width=2.0in]{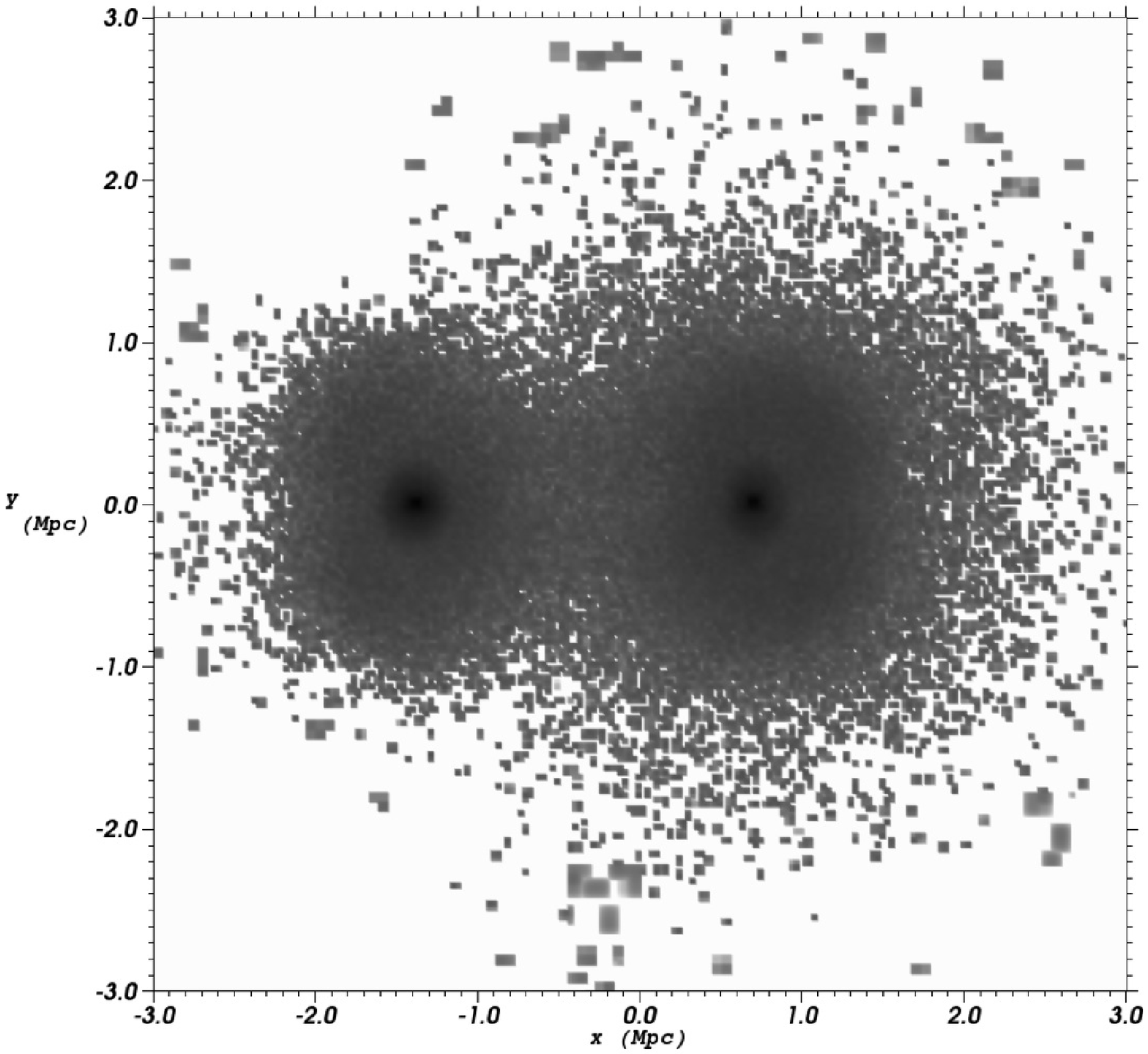}
\includegraphics[width=2.0in]{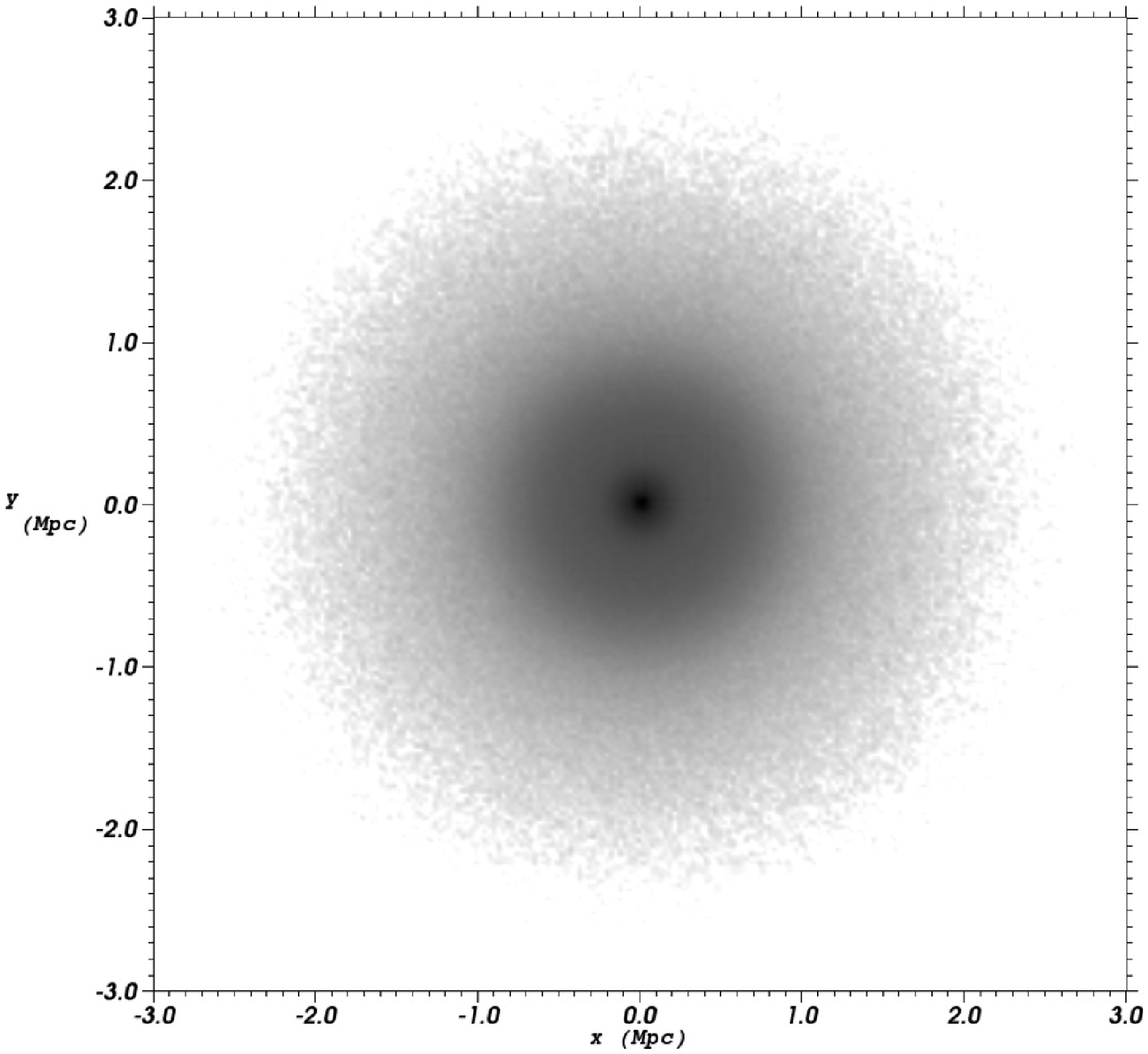}
\includegraphics[width=2.0in]{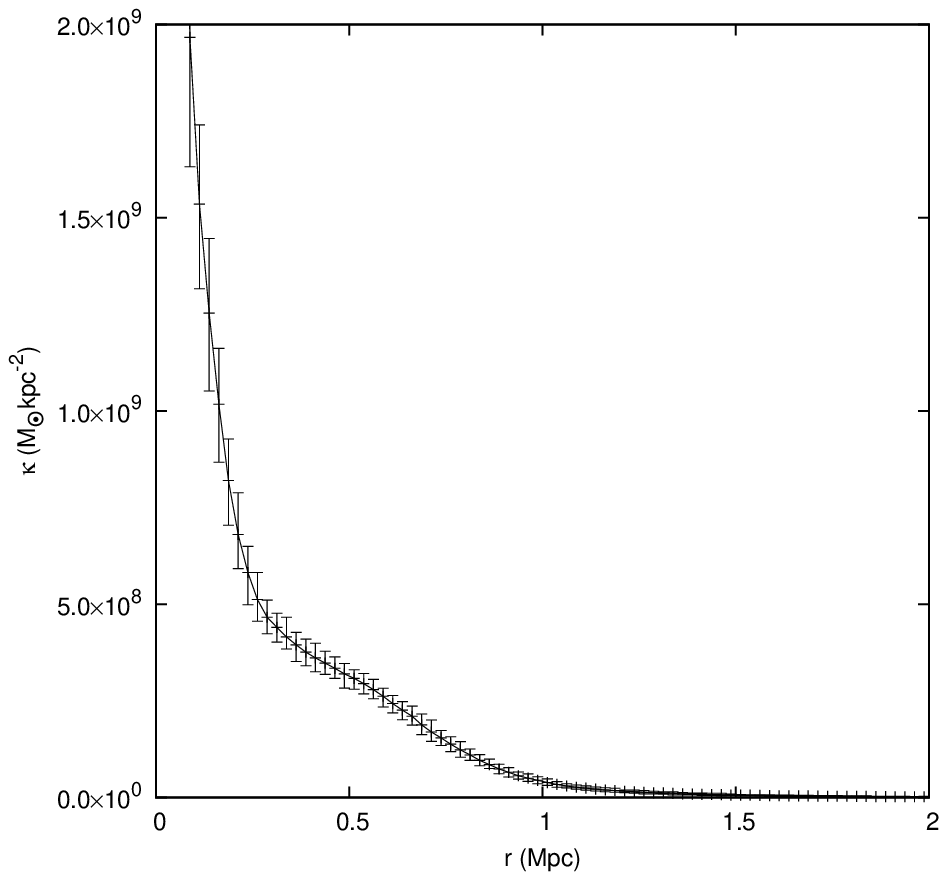}
\caption{$t$ = 1.5 Gyr snapshot of the $\beta$ = -3/2 simulation. Left: Slice through the $z = 0$ coordinate plane of dark matter density. Center: Dark matter density projected along the collision axis. Right: Projected density profile.\label{fig4}}
\end{figure*}

\begin{figure*}
\includegraphics[width=2.0in]{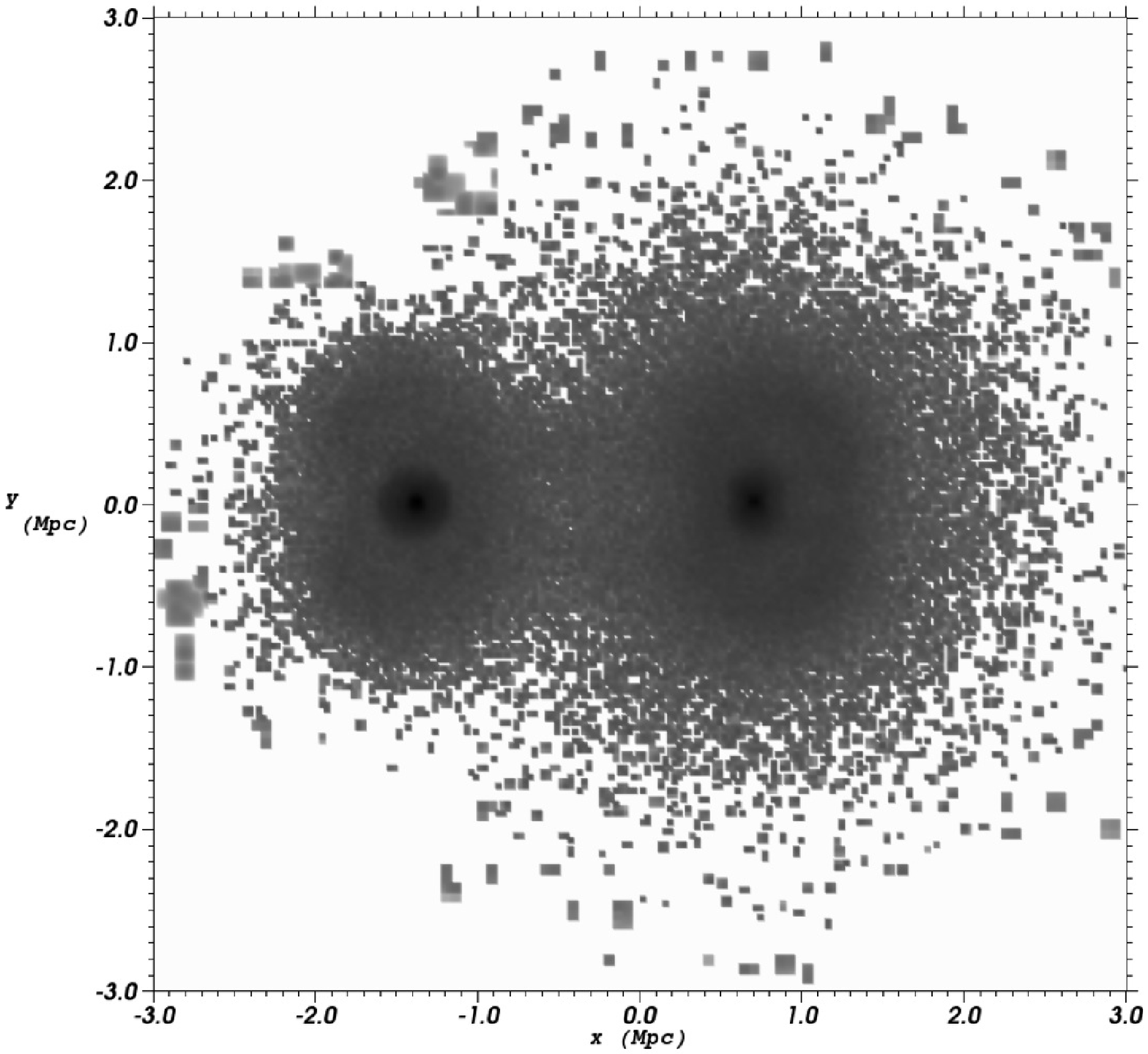}
\includegraphics[width=2.0in]{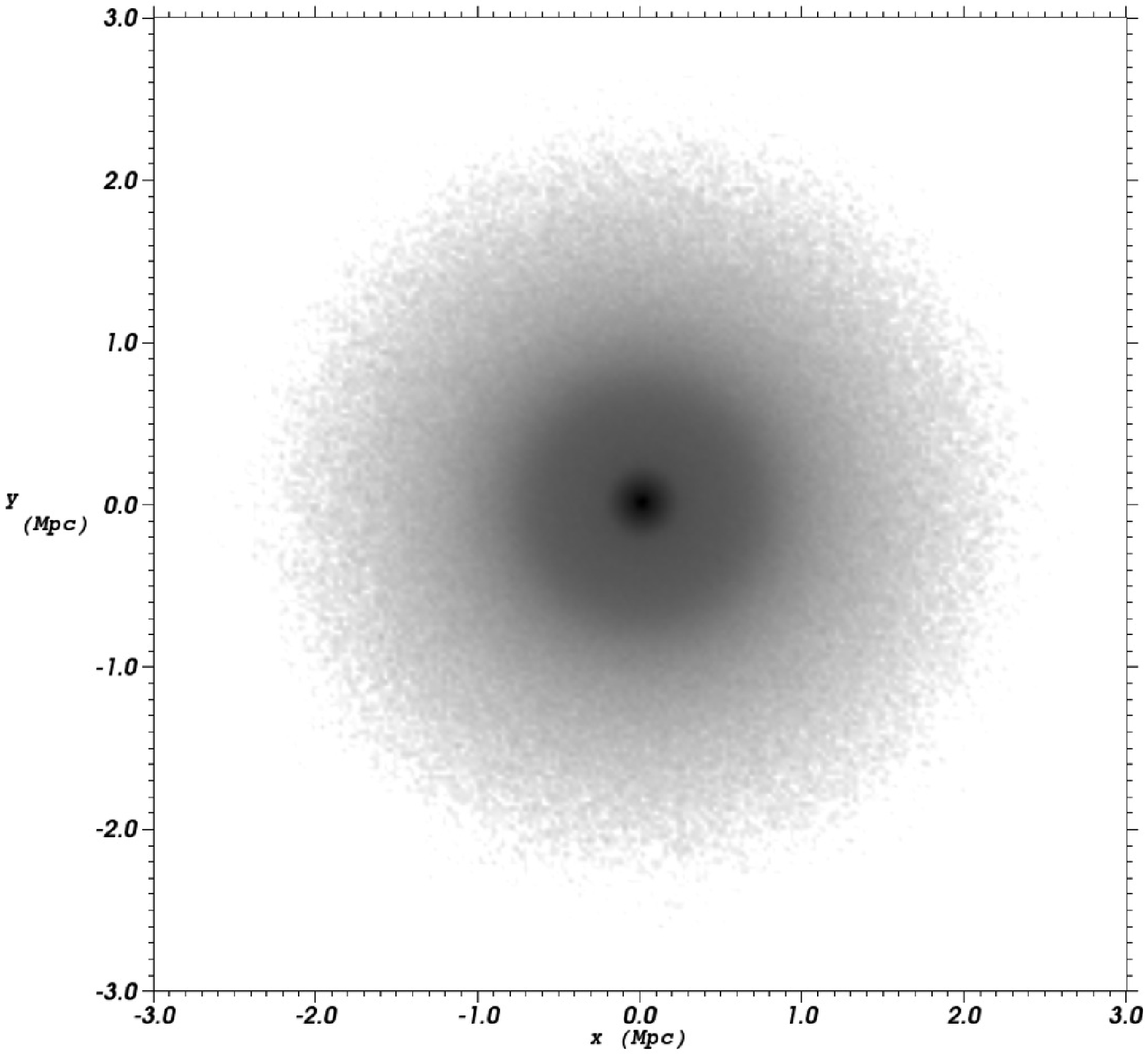}
\includegraphics[width=2.0in]{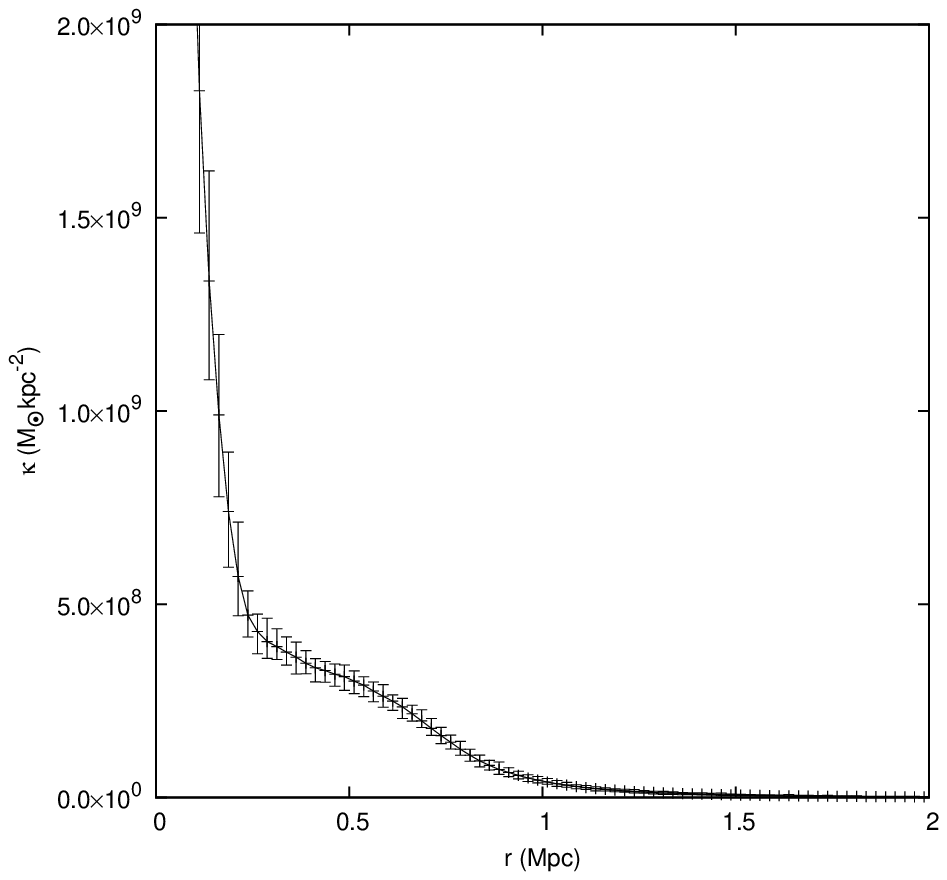}
\caption{$t$ = 1.5 Gyr snapshot of the $\beta$ = -5/2 simulation. Left: Slice through the $z = 0$ coordinate plane of dark matter density. Center: Dark matter density projected along the collision axis. Right: Projected density profile.\label{fig5}}
\end{figure*}

\begin{figure*}
\includegraphics[width=2.0in]{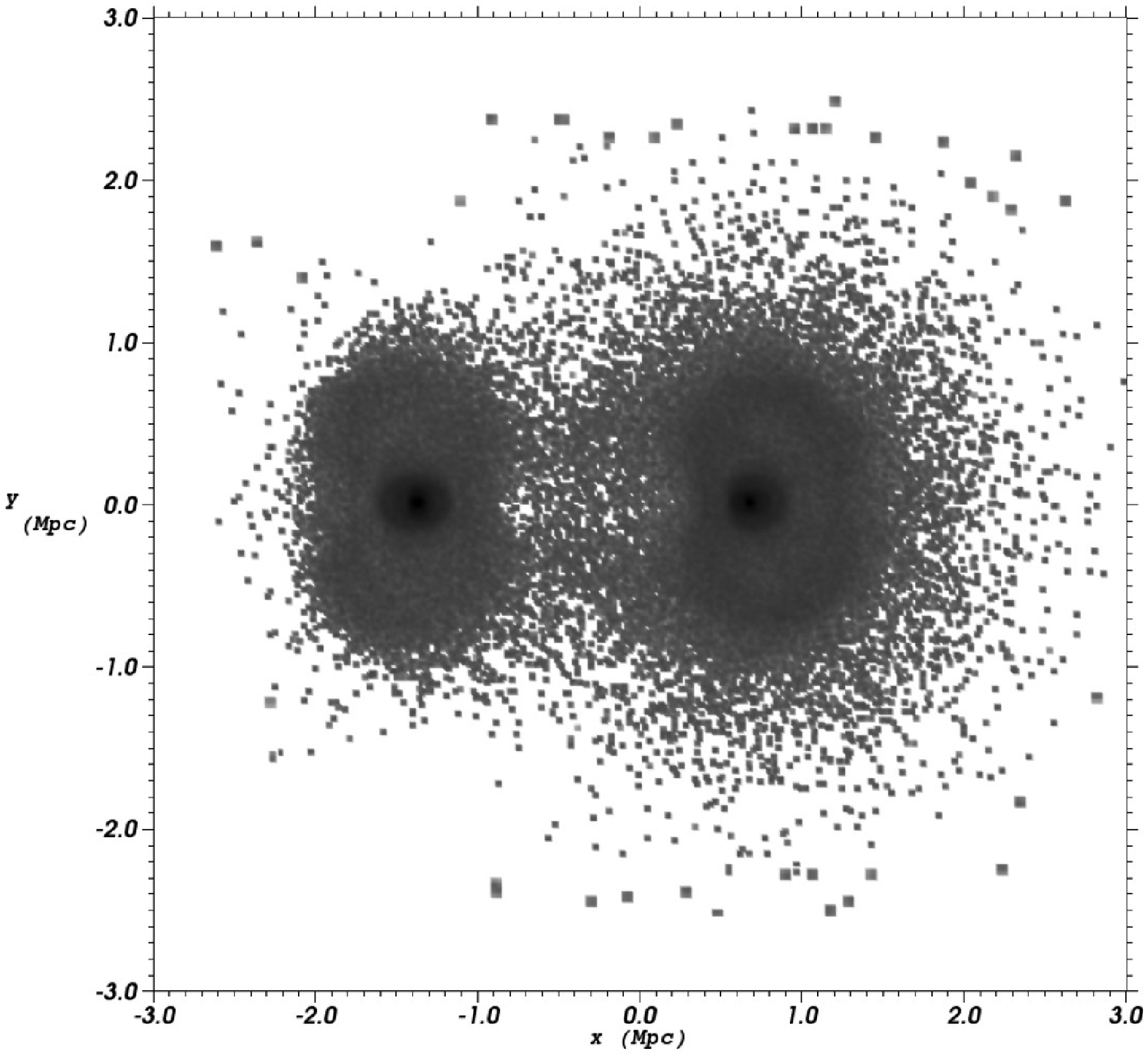}
\includegraphics[width=2.0in]{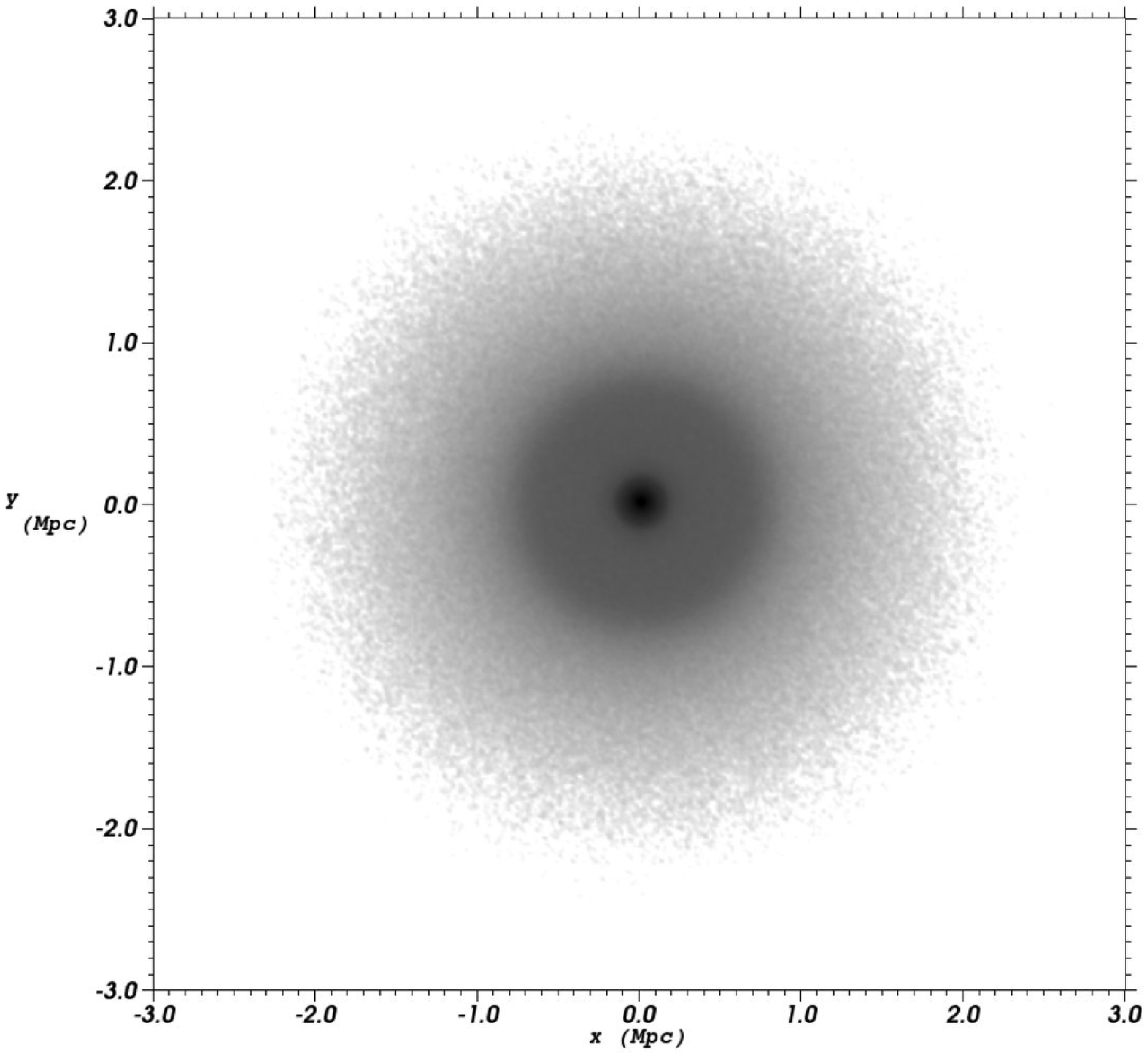}
\includegraphics[width=2.0in]{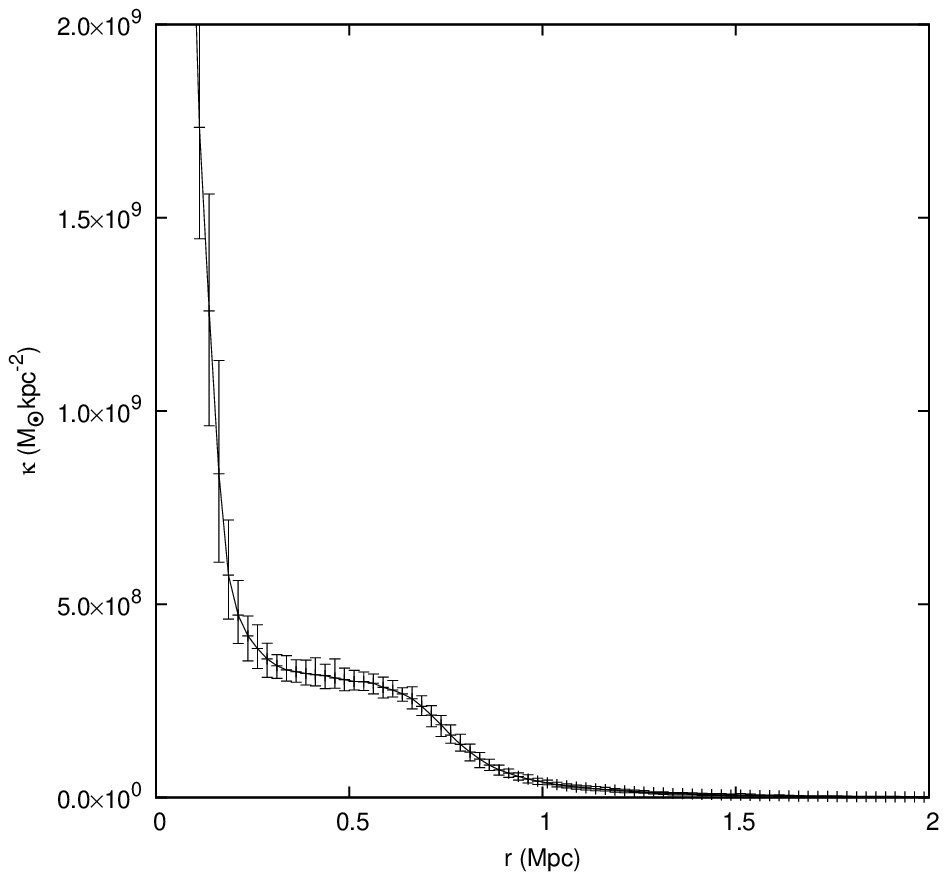}
\caption{$t$ = 1.5 Gyr snapshot of the $\beta$ = -17/2 simulation. Left: Slice through the $z = 0$ coordinate plane of dark matter density. Center: Dark matter density projected along the collision axis. Right: Projected densityprofile.\label{fig6}}
\end{figure*}

\begin{figure*}
\includegraphics[width=2.0in]{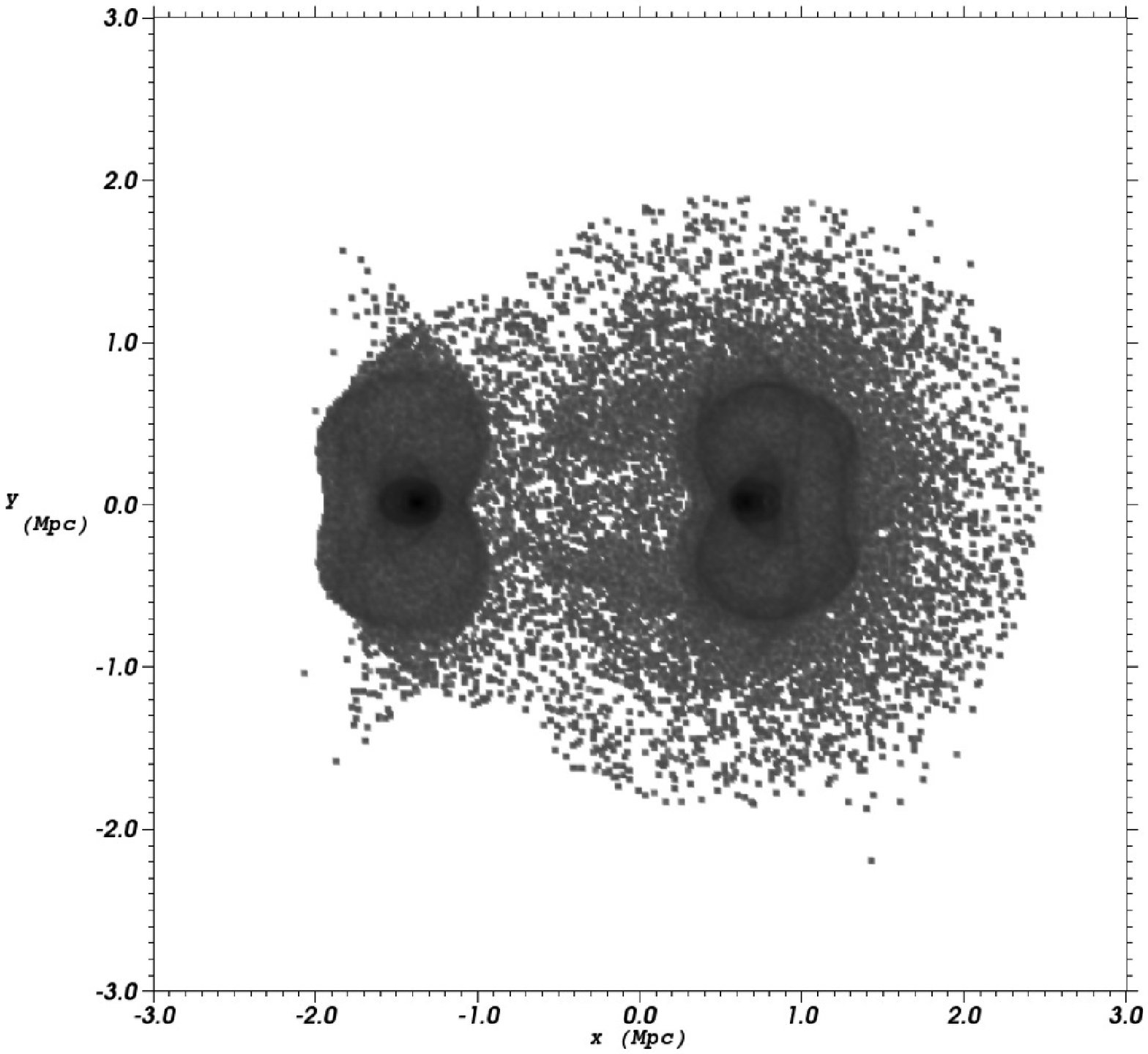}
\includegraphics[width=2.0in]{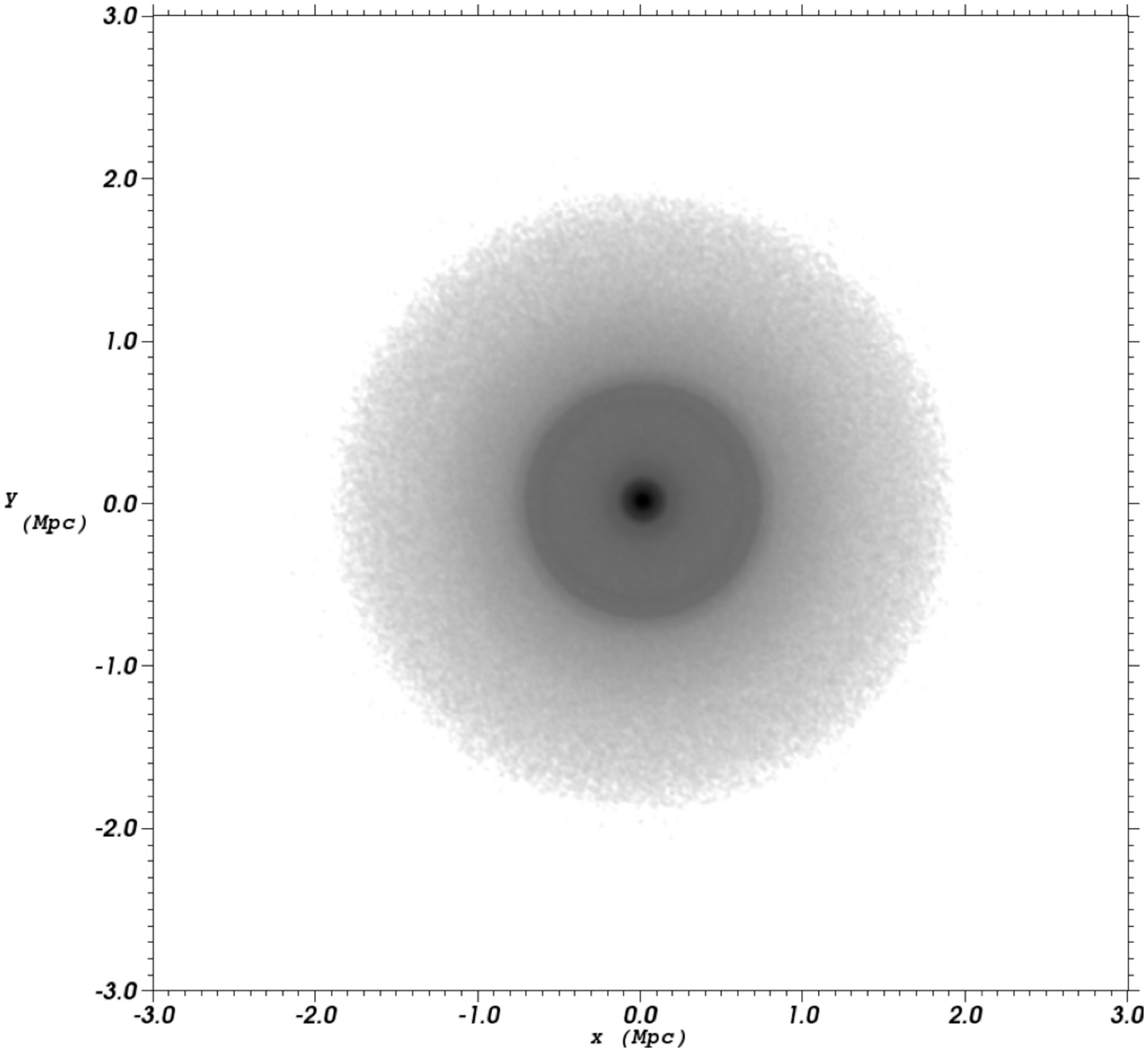}
\includegraphics[width=2.0in]{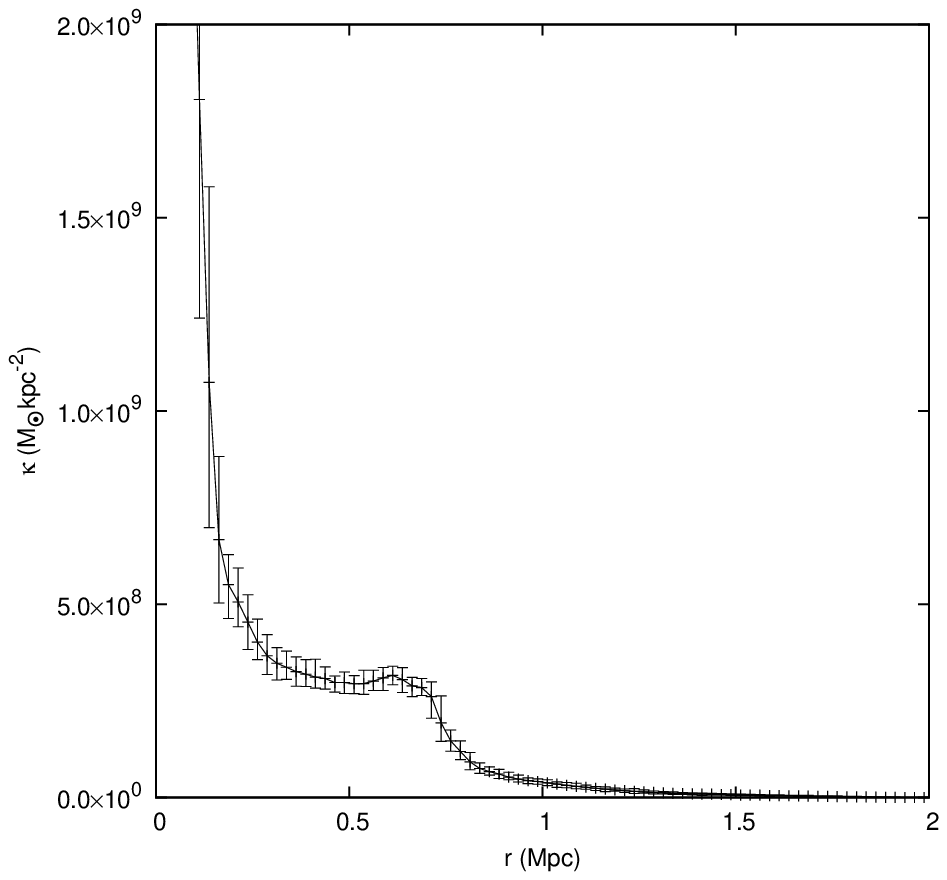}
\caption{$t$ = 1.5 Gyr snapshot of the $\beta = -\infty$ simulation. Left: Slice through the $z = 0$ coordinate plane of dark matter density. Center: Dark matter density projected along the collision axis. Right: Projected density profile.\label{fig7}}
\end{figure*}

\begin{figure*}[htbp]
\begin{center}
\epsscale{1.1}\plottwo{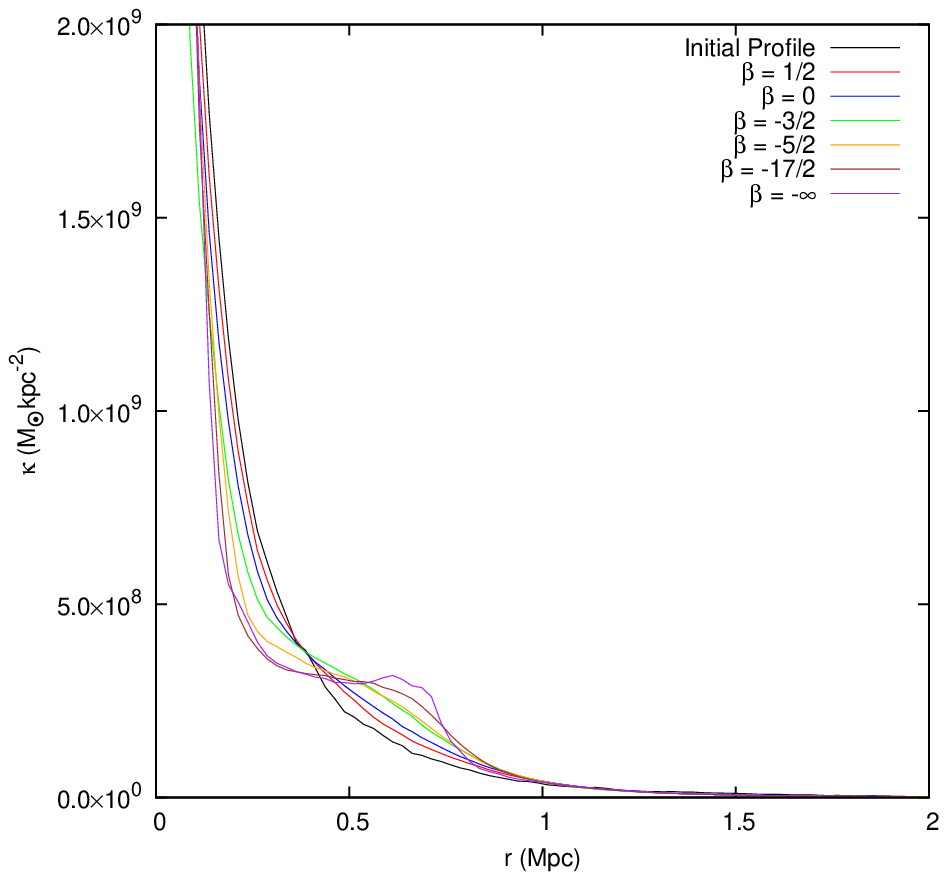}{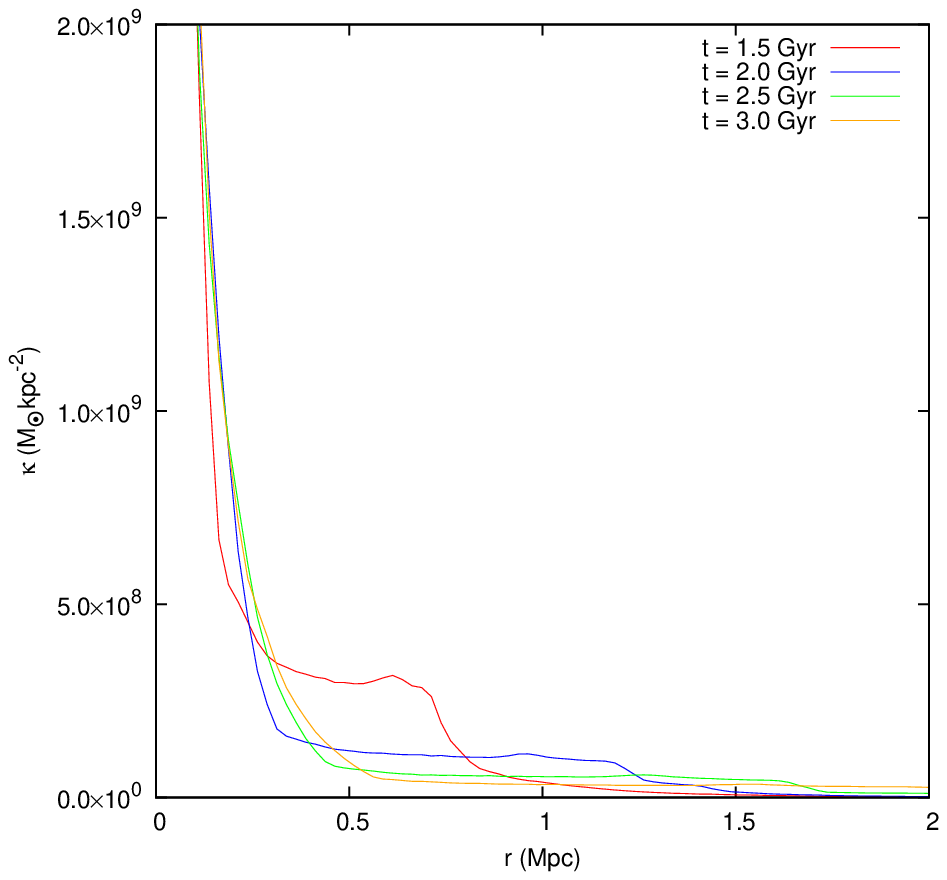}
\caption{Left: Angle-averaged DM density profiles at $t$ = 1.5 Gyr into the simulation ($\sim$0.6 Gyr post-collision) for varying $\beta$. As $\beta$ is made more negative a more pronounced shoulder-like feature appears in the density profile but a ring does not appear except in the $\beta = -\infty$ case. Right: Angle-averaged DM density profiles at the epochs $t$ = 1.5, 2.0, 2.5, and 3.0 Gyr for the $\beta = -\infty$ simulation. Note that the ringlike feature appears prominently at $t$ = 1.5 Gyr ($\sim$0.6 Gyr after the collision) but is a transient feature that has completely disappeared at later times.\label{fig8}}
\end{center}
\end{figure*}

Figures \ref{fig2} through \ref{fig7} show slices though the $z = 0$ coordinate plane of the dark matter density, the projected dark matter density, and the projected density profile for each simulation at the epoch $t = 1.5$ Gyr ($\approx 0.6$ Gyr after the collision, defined as the time when the centers of the halos are coincident). At this epoch, our simulations show no ring feature (defined as a "bump" in the radial distribution of dark matter particles projected onto the sky; i.e. an increase followed by a decrease in this distribution) for initial velocity distributions that are radially anisotropic ($\beta = 0.5$) and isotropic orbits ($\beta = 0$).  For initial velocity
distributions that are increasing tangentially anisotropic
($\beta = -0.5, -2.5, -8.5$), our simulations show a more pronounced shoulder but no ring.  Only for an initially circular velocity distribution ($\beta = - \infty$) do our simulations show a ring. Figure \ref{fig8} shows the angle-averaged projected density profile at $t = 1.5$ Gyr for all of the different initial values of $\beta$. The initial profile is also shown for comparison.

Even in the simulation in which our simulations show a ring ($\beta =
-\infty$), the feature is short-lived. Figure \ref{fig8} shows that the feature is prominent at epochs $t = 1.5$ Gyr and $t = 2.0$ Gyr
($\approx 0.6$ and $\approx 0.95$ Gyr after the collision).  However,
by $t = 2.5$ Gyr and $t = 3.0$ Gyr ($\approx 1.6$ and $\approx 2.1$ Gyr after the collision), it has become insignificant.


\section{Discussion}
\label{Sec:discussion}

\subsection{Relevance to Observations of Cl 0024+17}

Our simulations show that for physically and observationally motivated
dark matter particle distributions (NFW-like) and initial dark matter
velocity distributions with a wide range of angular anisotropies, no
ring feature forms as in the scenario of \citet{jee07}. The collision scenario for Cl 0024+17, as mentioned before, is supported by the results of \citet{czo02}, who demonstrated the existence of a bimodal redshift distribution in the cluster galaxies. They also showed that a simulation of such a collision can explain the observed flattening of the density profile in the cluster core even if the initial profiles are cuspy. However, they did not mention the existence of a ring feature in their paper.

\citet{jee07} presented results of a lensing analysis of Cl 0024+17.
In addition, they presented the results of simulations of head-on
collisions of galaxy cluster dark matter halos. The dark matter
density distributions of the galaxy clusters in their study were not
NFW-like with central cusps, but softened isothermal spheres with
$\rho(r) \propto [1+({r/r_{\rm c}})^2]^{-1}$; the initial velocity distribution of the particles was not reported. In these simulations, a clear ring-like feature is evident. 

\begin{figure*}
\includegraphics[width=2.0in]{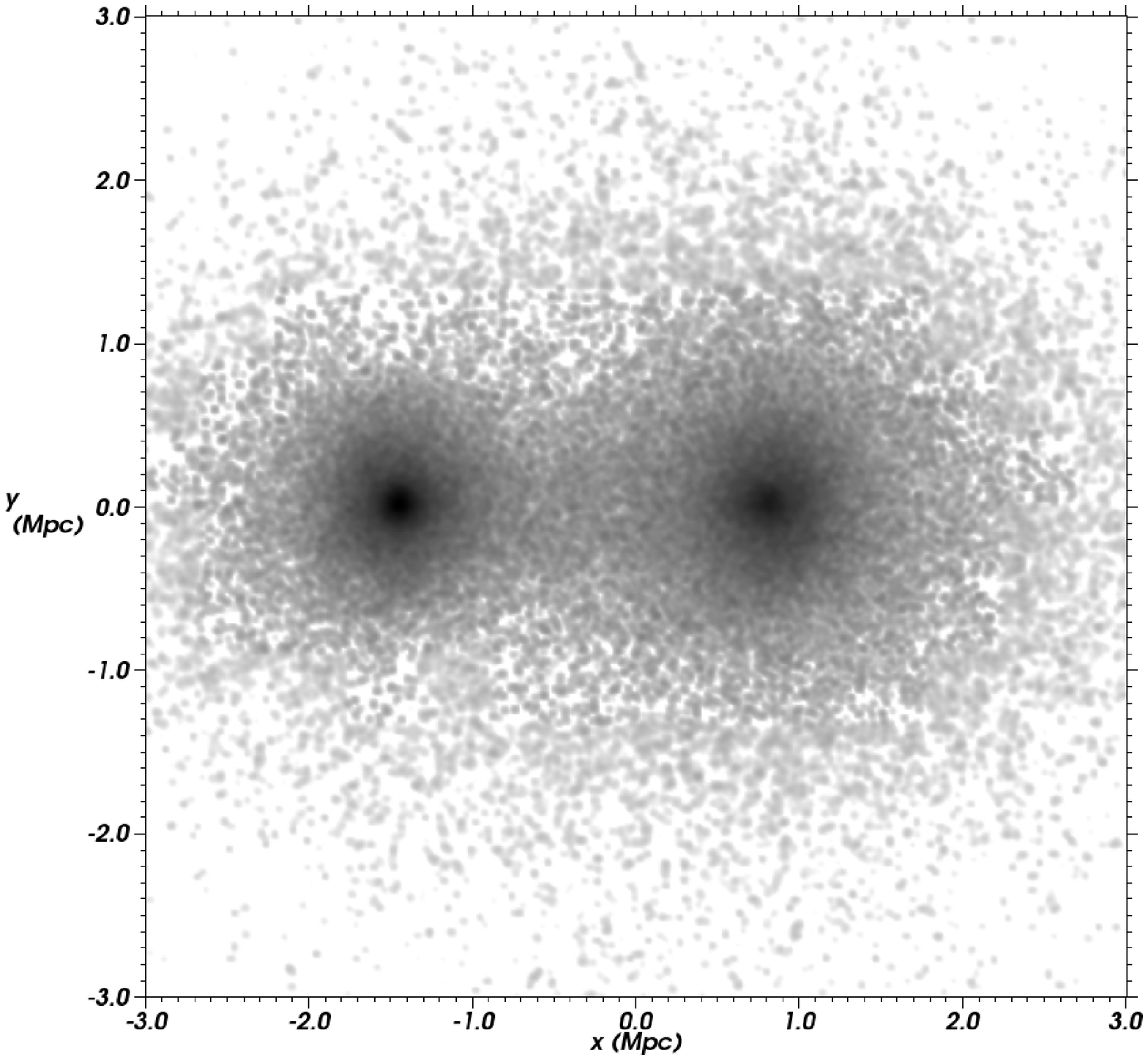}
\includegraphics[width=2.0in]{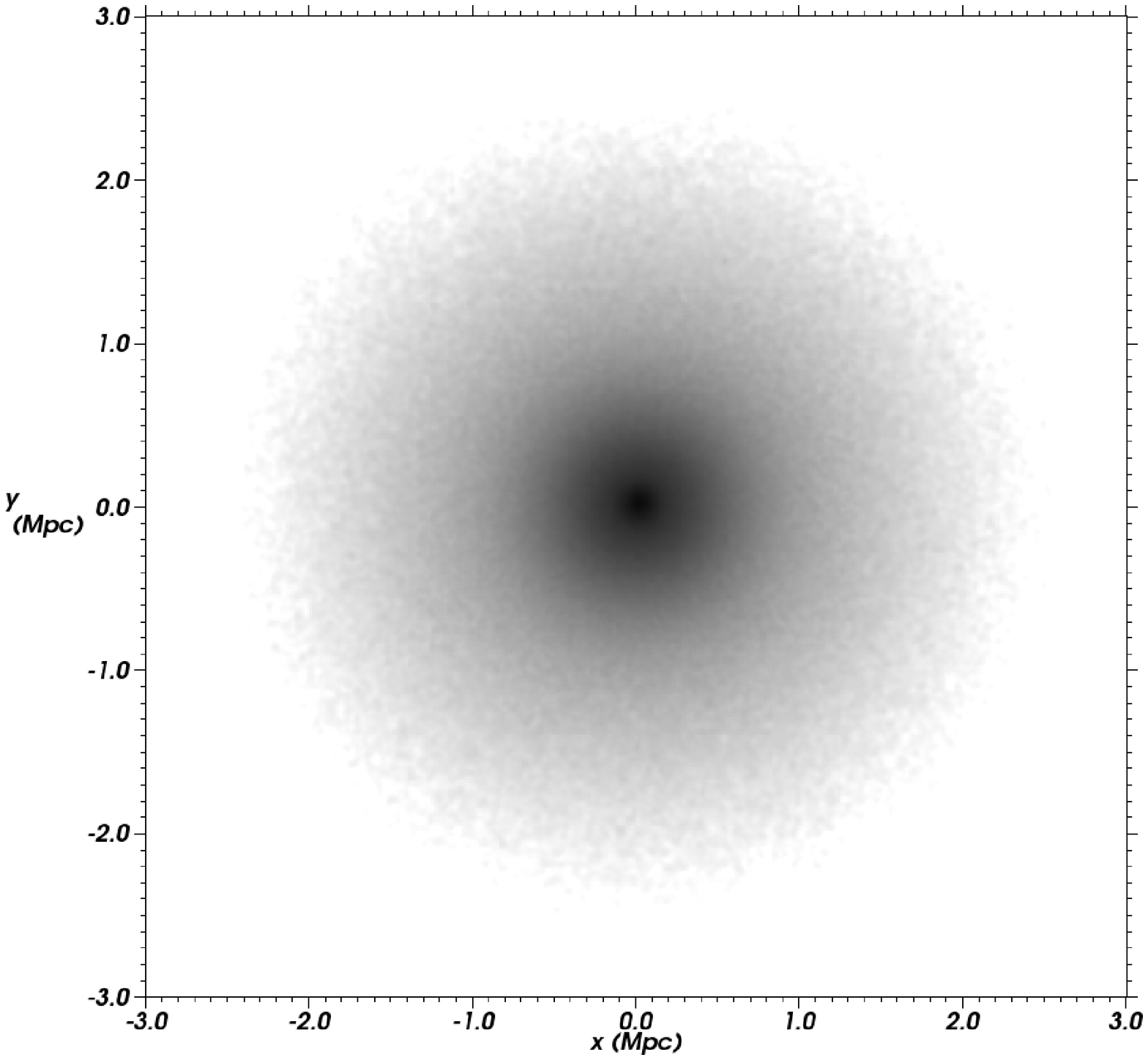}
\includegraphics[width=2.0in]{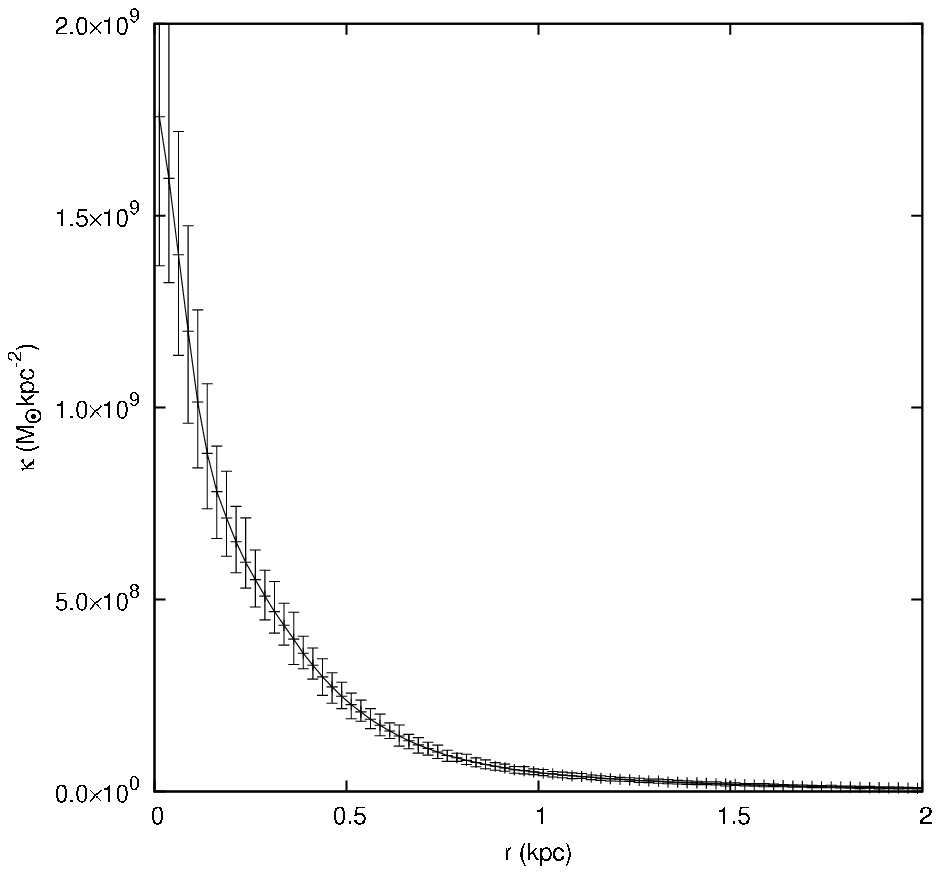}
\caption{$t$ = 1.5 Gyr snapshot of a cluster collision with softened isothermal sphere density profiles and $\beta$ = 0. Left: Slice through the $z = 0$ coordinate plane of dark matter density. Center: Dark matter density projected along the collision axis. Right: Projected density profile.\label{fig9}}
\end{figure*}

In an attempt to understand the difference between the results of our simulations and those of \citet{jee07}, we have performed a simulation assuming a softened isothermal sphere for the density profile with the same parameters, and the same cluster masses, central densities, and initial relative velocity as quoted in \citet{jee07}.  Unfortunately, jee et al 07 do not state the angular velocity distribution they use.  We assume an isotropic velocity distribution for the dark matter particles, in agreement with the velocity distribution in the inner regions of dark matter halos seen in $\Lambda$CDM simulations of structure formation.  Similarly to the $\beta$ = 0 case with the Hernquist profile, we find no evidence for a shoulder, let alone a ring-like feature (see \ref{fig9}). This shows, as one might expect, that a modest difference in the density profiles between the Hernquist profile and the softened isothermal sphere profile is not the origin of the difference between the two results.

Our simulations show no ring-like feature even for initial particle velocity distributions that are highly tangentially anisotropic; only for an initially circular velocity distribution do our simulations show a ring. As we discuss below, even modestly tangentially anisotropic velocity distributions are not expected for the dark matter particles in galaxy clusters. This, together with the results of our simulations imply that the production of ring-like features in galaxy cluster collisions is likely to be rare.


\subsection{Comparison with ``Ring Galaxies''}

The phenomenon of ``ring galaxies'' (e.g., the Cartwheel Galaxy) is explained by a similar mechanism, that of a smaller, more compact galaxy colliding with a disk galaxy \citep{lyn76}. $N$-body simulations of galaxy-galaxy collisions have demonstrated this phenomenon, for which the collisionless ``particles'' are stars \citep{thy77,too78}. In such simulations, the velocity distributions of the stars are set up to be tangentially anisotropic, or $\beta < 0$, and most of the orbits are circular. Also, the stars are concentrated into a disk-shaped structure rather than a spheroidal structure. \citet{lyn76} demonstrated that such a collision causes a crowding of particles on tangential orbits. In our simulations, as $\beta$ is made more negative a more pronunced shoulder appears in the projected dark matter density, but a ring appears only when the initial velocity distribution is circular. However, $\Lambda$CDM simulations of structure formation have demonstrated that in the inner regions of dark matter halos the particle velocities are nearly isotropic ($\beta \sim 0$), and they become more {\it radially} anisotropic in the outskirts of the halo \citep{col96}. Our results show that under such conditions no ring forms. If the formation of dark-matter ring features in collisions between galaxy clusters is dependent upon tangentially anisotropic velocities of the particles, then it would require fine-tuned, high-angular momentum initial conditions in the dark matter distribution, which is unrealistic under the standard CDM paradigm. 

\section{Conclusions}

A recent lensing analysis of the cluster Cl 0024+17 revealed possible evidence of a ring-like dark matter structure. This has been interpreted as the result of a collision between two clusters of galaxies along the line of sight, a scenario for which there are other lines of evidence in this system. We have performed simulations of collisions between galaxy cluster dark matter halos to test this hypothesis, using density profiles for the dark matter halos motivated by observations and simulations and investigating a parameter space of initial velocity distribution where we vary the initial velocity anisotropy. Our simulations show that a more pronounced ``shoulder'' feature appears in the projected dark matter density when we make the velocities of the dark matter particles more tangentially anisotropic, by analogy with the phenomena of "ring galaxies."  Our simulations show that, although the collision ejects a large amount of mass from the dark matter cores, a ring-like features does not form, even for initial velocity distributions that are highly tangentially anisotropic.  Only when the initial velocity distribution is circular does a ring form.  Since we do not expect dark matter particles in clusters of galaxies to have tangentially anisotropic velocities, our investigation of this parameter space leaves us without an explanation for the ring-like feature that appears in the dark matter distribution of Cl~0024+17.

\acknowledgments

This work is supported at the University of Chicago by the U.S Department of Energy (DOE) under Contract B523820 to the ASC Alliances Center for Astrophysical Nuclear Flashes. Calculations were performed using the computational resources of Lawrence Livermore National Laboratory and Los Alamos National Laboratory. JAZ is grateful to Andrey Kravtsov and Douglas Rudd for useful discussions and advice, and to James Jee for information about their study of Cl~0024+17. JAZ is supported by the Department of Energy Computational Science Graduate Fellowship, which is provided under grant number DE-FG02-97ER25308.

\end{document}

%% file: tab1.tex
\begin{table}[thdp]
\caption{Velocity Anisotropy\label{tab1}}
\begin{center}
\begin{tabular}{cccc}
\hline
\hline
Simulation & $\beta$ & $\sigma_\theta/\sigma_r$ & $E$ (erg) \\
\hline
S1 & 1/2 & 0.707 & $-4.45 \times 10^{63}$ \\
S2 & 0 & 1 & $-4.38 \times 10^{63}$ \\ 
S3 & -3/2 & 1.58 & $-4.08 \times 10^{63}$ \\
S4 & -5/2 & 1.87 & $-4.98 \times 10^{63}$ \\
S5 & -17/2 & 3.08 & $-5.28 \times 10^{63}$ \\
S6 & $-\infty$ & $\infty$ & $-4.88 \times 10^{63}$ \\
\hline
\end{tabular}
\end{center}
\end{table}

%% file: tab2.tex
\begin{table}[thdp]
\caption{Initial Cluster Parameters\label{tab2}}
\begin{center}
\begin{tabular}{ccccc}
\hline
\hline
Cluster & $M_0$ (M$_{\sun}$) & $R$ ({\rm kpc}) & $a$ ({\rm kpc}) & $\rho_s$ ($M_\odot kpc^{-3}$) \\
\hline
1 & $6.0 \times 10^{14}$ & 2000.0 & 400.0 & $1.49 \times 10^6$ \\
2 & $3.0 \times 10^{14}$ & 1000.0 & 200.0 & $5.97 \times 10^6$ \\
\hline
\end{tabular}
\end{center}
\end{table}